\documentclass[aps,prd,preprint,groupedaddress]{revtex4-1}

\usepackage{bm}
\usepackage{slashed}
\usepackage{url}
\usepackage{float}

\usepackage{amsmath}
\usepackage{graphicx}
\usepackage{comment}
\usepackage{subfigure}
\usepackage{float}

\newcommand{\be}{\begin{equation}}
\newcommand{\ee}{\end{equation}}
\newcommand{\bea}{\begin{eqnarray}}
\newcommand{\eea}{\end{eqnarray}}

\begin{document}

\title{Ground state of Yang-Mills theory in 2+1 dimensions}

\author{Marco Frasca}\email{marcofrasca@mclink.it}
\affiliation{via Erasmo Gattamelata, 3, 00176 Roma, Italy}

\begin{abstract}
Yang-Mills theory in 2+1 dimensions showed to be a research area yielding firm results in theoretical physics when compared to lattice computations. Recent analysis displayed astonishing agreement for the value of the string tension and excellent comparison for the spectrum. This successful approach can be put at test with a different theoretical framework that we devised in our preceding work for the scalar field theory in the strong coupling limit. The confirmations we get are really striking supporting it in full. As a by-product we are also able to show how AdS/CFT approach, with a description using flux tubes, is supported exactly as expected in the Isgur-Paton model with a dimensionless correction factor for the ground state of the theory already determined in lattice computations.
\end{abstract}

\date{\hfill}

\maketitle

\section{Introduction}

A deep understanding of Yang-Mills theory in all the range of the coupling represents a fundamental aspect of our comprehension of strong interactions. The reason is that this would open the possibility to accomplish computations of the behavior of the theory in the low-energy limit where the theory displays bound states. Currently, the only way to obtain results that are derived directly from the theory is through extensive use of lattice computations on large computer facilities. This has permitted for the Yang-Mills theory to obtain both the spectrum and the behavior of propagators in several gauges also at finite temperature \cite{Durr:2008zz,Bazavov:2009bb,Maas:2011se,Petreczky:2012rq} in four dimensions and similarly for the case $d=2+1$ \cite{Teper:1998te,Lucini:2002wg,Bringoltz:2006zg,Caselle:2011fy,Caselle:2011mn}. Specially in this latter case, very precise results exist for the string tension and the spectrum.

From a theoretical standpoint, the situation appears decisively better for the three-dimensional case where some analisys have been performed producing excellent agreement with lattice computations for the string tension \cite{Nair:2002yg,Karabali:2009rg} and the spectrum \cite{Leigh:2005dg,Leigh:2006vg}. The starting point was a work by Karabali and Nair that proposed a proper set of matrix variables to work with in this case to put forward a wavefunction and derive fundamental results of the theory \cite{Karabali:1995ps}. Karabali and Nair approach appears greatly successful in the derivation of the string tension and higher order corrections \cite{Karabali:2009rg}. For the spectrum, a different wavefunction was postulated \cite{Leigh:2005dg,Leigh:2006vg} always in the framework of Karabali and Nair formalism. Again, the agreement with lattice data was impressive.

In a recent paper of ours we were able to build a quantum field theory for the self-interacting scalar field in the limit of the coupling running to infinity \cite{Frasca:2013tma}. We displayed a set of classical solutions that, notwithstanding we started from a massless equation, showed a massive dispersion relation. These solutions were already proposed in \cite{Frasca:2009bc} but the idea in \cite{Frasca:2013tma} was to consider them as the vacuum expectation value of the field and build the quantum theory from them. In this way one has that conformal invariance is broken and a zero mode appears. The particles get a mass and a tower of excited states represented described by the spectrum of a harmonic oscillator. This theory shares a trivial infrared trivial fixed point and an ultraviolet trivial fixed point making the theory overall trivial but with a mass gap. On this ground it is a natural question to ask if also a Yang-Mills theory can share such classical solutions and a corresponding quantum field theory built upon them. The answer was affirmative as we showed in \cite{Frasca:2009yp} but this is true asymptotically in the general case while the resul holds exactly just in the Lorenz (Landau) gauge. The corresponding quantum field theory develops a mass gap but is trivial at both sides of the range due to the trivial infrared fixed point and asymptotic freedom on the other side.

In this paper we develop this approach alternative to the Nair and Karabali formalism with the only idea in mind to validate it. That is, we aim to try to solve the Yang-Mills theory in 2+1 dimensions in another way and we will show how successful indeed is the Karabali and Nair technique. As a by-product we will get an important hint that the flux tube description of hadron emerging in AdS/CFT approach \cite{Brodsky:2014yha} is a successful one as we are able to get the right ground state of the theory by introducing the same factor as in the Isgur-Paton theory \cite{Isgur:1984bm} as demonstrated by Teper and Johnson \cite{Johnson:2000qz}. The agreement is exceedingly good so that our error in the value of the ground state is significantly improved with respect to the works \cite{Leigh:2005dg,Leigh:2006vg}.

The paper is so structured. In Sec.~\ref{sec1} we introduce the classical solutions. The gluon propagator is discussed in Sec.~\ref{sec2}. We provide the formalism for higher order corrections in Sec.~\ref{sec3}. Numerical comparisons are presented in Sec.~\ref{sec4}. Finally, conclusions are given in Sec.~\ref{sec5}.

\section {Classical solutions}
\label{sec1}

\subsection{General case}
\label{ssec21}

Motion equation for Yang-Mills theory can be straightforwardly written down for any number of dimensions and SU(N) group in the form
\begin{equation}
\label{eq:YM}
    {\cal D}^\mu F_{\mu\nu}=0
\end{equation}
being
\begin{equation}
   {\cal D}_\mu=\partial_\mu-igT^aA^a_\mu
\end{equation}
the covariant derivative, $T_a$ the generators of the group and $A^a_\mu$ the potentials ($a,b,c,\ldots$ are color indexes running from 1 to N), and
\begin{equation}
    F_{\mu \nu}^a = \partial_\mu A_\nu^a-\partial_\nu A_\mu^a+gf^{abc}A_\mu^bA_\nu^c
\end{equation}
the field components with $F_{\mu\nu}=T^aF^a_{\mu\nu}$ and $f^{abc}$ the structure constants of the group. As our aim is to work out a result in quantum field theory, we add a term into eq.(\ref{eq:YM}) to fix the gauge in the form
\begin{equation}
   -\left(1-\frac{1}{\xi}\right)\partial_\nu(\partial\cdot A^a)
\end{equation}
with $\xi$ a free parameter determining the gauge choice.

Using perturbation theory, one can show that there exists a set of solutions of Yang-Mills equations of motion that can be cast in the form \cite{Frasca:2009yp}
\begin{equation}
\label{eq:asymp}
   A_\mu^a(x)=\eta_\mu^a\chi(x)+O\left(1/Ng^2\right).
\end{equation}
This holds provided
\begin{equation}
\label{eq:scal}
   \partial^2\chi(x)+Ng^2\chi(x)=0
\end{equation}
and $\eta_\mu^a$ are some constants to be determined depending on the problem at hand. These solutions become exact and not just perturbative for the Lorenz (Landau) gauge. An interesting aspect of these solutions is that hold in any dimensions $d>2$. For $d=2$ Yang-Mills equations of motion are trivial and no such solutions can be found.

Without exploiting all the possible solutions of eq.(\ref{eq:scal}) we limit our interest to a subclass of solutions that have the property to be massive even if we started from massless equations of motion. We have fully exploited this case in Ref.\cite{Frasca:2013tma}. In this paper we consider as a ground state of the quantum field theory of a scalar field such exact solutions. In 3+1 dimensions this can be written down as \cite{Frasca:2013tma}
\begin{equation}
    \chi_{d=3+1}(x)=\mu\left(\frac{2}{Ng^2}\right)^\frac{1}{4}\operatorname{sn}\left(k\cdot x+\phi,-1\right)
\end{equation}
being sn a Jacobi elliptic function, $\phi$ an arbitrary phase, $\mu$ an arbitrary constant having the dimension of a mass and provided that
\begin{equation}
    k^2=\sqrt{\frac{Ng^2}{2}}\mu^2.
\end{equation}
So, if we interpret $k$ as a four-vector of momenta, this can be seen as the dispersion relation of a massive wave. These solutions are rather counterintuitive as we started from a pure massless theory. A mass term can be seen to arise from the nonlinearities of the equations we started from. In the following we will assume that such solutions are just the ground state for the quantum field theory we aim to study.

Being $Ng^2$ dimensionless, one can always use this constant to properly rescale all the physical variables to get
\begin{equation}
    \chi_{d=3+1}(x)=\mu 2^\frac{1}{4}\operatorname{sn}\left(k\cdot x+\phi,-1\right)
\end{equation}
being sn a Jacobi elliptic function, $\phi$ an arbitrary phase, $\mu$ an arbitrary constant having the dimension of a mass and provided that
\begin{equation}
    k^2=\frac{Ng^2}{\sqrt{2}}\mu^2.
\end{equation}

\subsection{Case $d=2+1$}
\label{ssec22}

In 2+1 dimensions Yang-Mills equations have a coupling $g^2$ having the dimension of a mass or inverse of a length. This means that our solution takes the form
\begin{equation}
\label{eq:exsol}
    \chi_{d=2+1}(x)=2^\frac{1}{4}\sqrt{Ng^2}\operatorname{sn}\left(k\cdot x+\phi,-1\right)
\end{equation}
and now one has
\begin{equation}
\label{eq:ds2+1}
    k^2=\frac{N^2g^4}{\sqrt{2}}.
\end{equation}

%\subsection{Case d>4}
%
%For dimensions greater than 4 we note that $D_{Ng^2}=L^{d-4}$ and $D_\chi=L^{-\frac{d-2}{2}}$. This yields

\section{Gluon propagator}
\label{sec2}

We need to introduce the propagator of Yang-Mills theory in the infrared limit. This is generally accomplished by a current expansion \cite{Frasca:2013tma,Cahill:1985mh}. Instead to start from the action, we prefer the equations of motion \cite{Rubakov:2002fi}
\begin{eqnarray}
&&\partial^\mu\partial_\mu A^a_\nu-\left(1-\frac{1}{\alpha}\right)\partial_\nu(\partial^\mu A^a_\mu)+gf^{abc}A^{b\mu}(\partial_\mu A^c_\nu-\partial_\nu A^c_\mu)+gf^{abc}\partial^\mu(A^b_\mu A^c_\nu) \\ \nonumber
&&+g^2f^{abc}f^{cde}A^{b\mu}A^d_\mu A^e_\nu = j^a_\nu.
\end{eqnarray}
Then, we assume a functional form $A^a_\nu=A^a_\nu[j]$ and perform a Taylor expansion around the asymptotic solution (\ref{eq:asymp}). We have to take in mind that, for the Landau gauge, these solutions are exact but just asymptotic for whatever other gauge choice. So, we take in general
\begin{equation}
   A^a_\nu[j(x)]=\eta_\nu^a\chi(x)+\int d^dx'\left.\frac{\delta A_\nu^a}{\delta j_\mu^b(x')}\right|_{j=0}j_\mu^b(x')+
	\frac{1}{2}\int d^dx'd^dx''\left.\frac{\delta^2 A_\nu^a}{\delta j_\mu^b(x')\delta j_\kappa^c(x'')}\right|_{j=0}
	j_\mu^b(x')j_\kappa^c(x'')+\ldots.
\end{equation}
We are assuming here that eq.(\ref{eq:asymp}) represents the ground state of the theory i.e. $A^a_\nu[0]=\eta_\nu^a\chi(x)$. These describe oscillations around a vacuum expectation value of the fields as seen from our solutions in Sec.~\ref{ssec21} and \ref{ssec22}. Then, the propagator of the theory will be
\begin{equation}
   G_{\mu\nu}^{ab}(x,x')=\left.\frac{\delta A_\nu^a(x)}{\delta j_\mu^b(x')}\right|_{j=0}.
\end{equation}
We can obtain the corresponding equation by doing the functional derivative on the equation of motion. We get
\begin{eqnarray}
&&\partial^2 \frac{\delta A_\nu^a(x)}{\delta j_\rho^e(x')}
-\left(1-\frac{1}{\alpha}\right)\partial_\nu\left(\partial^\mu\frac{\delta A_\mu^a(x)}{\delta j_\rho^e(x')}\right) \nonumber \\
&&+gf^{abc}\frac{\delta A_\mu^b(x)}{\delta j_\rho^e(x')}\left(\partial^\mu A^c_\nu-\partial_\nu A^{\mu c}\right) \nonumber \\
&&+gf^{abc}A_\mu^b\left(\partial^\mu\frac{\delta A^c_\nu(x)}{\delta j_\rho^e(x')}
-\partial_\nu\frac{\delta A^{\mu c}(x)}{\delta j_\rho^e(x')}\right) \nonumber \\
&&+gf^{abc}\partial^\mu\left(\frac{\delta A_\mu^b(x)}{\delta j_\rho^e(x')} A^c_\nu\right) 
+gf^{abc}\partial^\mu\left(A_\mu^b\frac{\delta A_\nu^c(x)}{\delta j_\rho^e(x')}\right) \\ \nonumber
&&+g^2f^{abc}f^{cdh}\frac{\delta A_\mu^b(x)}{\delta j_\rho^e(x')}A^d_\mu A^h_\nu \\ \nonumber
&&+g^2f^{abc}f^{cdh}A^{b\mu}\frac{\delta A_\mu^d(x)}{\delta j_\rho^e(x')} A^h_\nu \\ \nonumber
&&+g^2f^{abc}f^{cdh}A^{b\mu}A^d_\mu\frac{\delta A_\nu^h(x)}{\delta j_\rho^e(x')}
= \delta_{ae}\eta_{\nu\rho}\delta^d(x-x').
\end{eqnarray}
Imposing $j=0$ one obtains the following equation for the Green function of Yang-Mills theory
\begin{eqnarray}
&&\partial^2G_{\nu\rho}^{ae}(x,x')
-\left(1-\frac{1}{\alpha}\right)\partial_\nu\partial^\mu G_{\mu\rho}^{ae}(x,x') \nonumber \\
&&+gf^{abc}G_{\mu\rho}^{be}(x,x')\left(\partial^\mu A^c_\nu-\partial_\nu A^{\mu c}(x)\right) \nonumber \\
&&+gf^{abc}A_\mu^b\left(\partial^\mu G_{\nu\rho}^{ce}(x,x')
-\partial_\nu G_{\mu\rho}^{ce}(x,x')\right) \nonumber \\
&&+gf^{abc}\partial^\mu \left(A^c_\nu G_{\mu\rho}^{be}(x,x')\right)
+gf^{abc}\partial^\mu\left(A_\mu^b G_{\nu\rho}^{ce}(x,x')\right) \\ \nonumber
&&+g^2f^{abc}f^{cdh}G_{\mu\rho}^{be}(x,x')A^{\mu d} A^h_\nu \\ \nonumber
&&+g^2f^{abc}f^{cdh}A^{b\mu}G_{\mu\rho}^{de}(x,x') A^h_\nu \\ \nonumber
&&+g^2f^{abc}f^{cdh}A^{b\mu}A^d_\mu G_{\nu\rho}^{he}(x,x')
= \delta_{ae}\eta_{\nu\rho}\delta^d(x-x').
\end{eqnarray}
or
\begin{eqnarray}
&&\partial^2G_{\nu\rho}^{ae}(x,x')
-\left(1-\frac{1}{\alpha}\right)\partial_\nu\partial^\mu G_{\mu\rho^{ae}}(x,x') \nonumber \\
&&+gf^{abc}G_{\mu\rho}^{be}(x,x')\left(\partial^\mu (\eta^c_\nu\chi(x))-\partial_\nu(\eta^{\mu c}\chi(x))\right) \nonumber \\
&&+gf^{abc}\eta_\mu^b\chi(x)\left(\partial^\mu G_{\nu\rho}^{ce}(x,x')
-\partial_\nu G_{\mu\rho}^{ce}(x,x')\right) \nonumber \\
&&+gf^{abc}\partial^\mu \left(\eta^c_\nu\chi(x) G_{\mu\rho}^{be}(x,x')\right)
+gf^{abc}\partial^\mu\left(\eta_\mu^b\chi(x) G_{\nu\rho}^{ce}(x,x')\right) \\ \nonumber
&&+g^2f^{abc}f^{cdh}G_{\mu\rho}^{be}(x,x')\eta^{\mu d} \eta^h_\nu\chi^2(x) \\ \nonumber
&&+g^2f^{abc}f^{cdh}\eta^{b\mu}G_{\mu\rho}^{de}(x,x') \eta^h_\nu\chi^2(x) \\ \nonumber
&&+g^2f^{abc}f^{cdh}\eta^{b\mu}\eta^d_\mu G_{\nu\rho}^{he}(x,x')\chi^2(x)
= \delta_{ae}\eta_{\nu\rho}\delta^d(x-x').
\end{eqnarray}
In order to compute the propagator, we perform a gauge's choice. The most common is the Landau gauge ($\alpha=1$) that also grants that we are using exact formulas rather than asymptotic ones. So, we write as usual for this gauge
\begin{equation}
  G_{\mu\nu}^{ab}(x,x')=\delta_{ab}\left(g_{\mu\nu}-\frac{p_\mu p_\nu}{p^2}\right)\Delta(x,x')
\end{equation}
being $p_\mu$ the momentum 4-vector. This yields for the above equation
\begin{equation}
   \partial^2\Delta(x,x')+3Ng^2\chi^2(x)\Delta(x,x')=\delta^d(x-x')
\end{equation}
that is the equation we were looking for. This equation coincides with that of the Green function of the scalar field obtained in \cite{Frasca:2013tma} in agreement with the mapping we derived in \cite{Frasca:2009yp} provided $\lambda\leftrightarrow Ng^2$, being $\lambda$ the corresponding coupling for the scalar field theory.

\subsection{Green function in $d=2+1$}

We now limit our analysis to the case $d=2+1$ and compute the exact Green function for this problem. The technique we follow is that outlined in Ref.\cite{Frasca:2013tma}. We just note that we have two independent solutions of the homogeneous equation
\begin{equation}
   \partial^2 y(x)+3Ng^2\chi^2(x)y(x)=0
\end{equation}
or
\begin{equation}
   \partial^2 y(x)+3\sqrt{2}(Ng^2)^2\operatorname{sn}^2\left(k\cdot x+\phi,-1\right)y(x)=0
\end{equation}
that are
\begin{equation}
\label{eq:sol1}
   y_1(t)={\rm cn}(p\cdot x+\phi,-1){\rm dn}(p\cdot x+\phi,-1)
\end{equation}
that holds provided
\begin{equation}
\label{eq:ds1}
    p^2=\frac{N^2g^4}{\sqrt{2}}.
\end{equation}
The other one can be obtained by writing it as
\begin{equation}
   y_2(x)=y_1(x)\cdot w(x)
\end{equation}
with
\begin{equation}
   {\rm cn}(p\cdot x+\phi,-1){\rm dn}(p\cdot x+\phi,-1)\partial^2 w-4{\rm sn}^3(p\cdot x+\phi,-1)p\cdot\partial w=0.
\end{equation}
Now, we introduce a new variable $\bar x=p\cdot x+\phi$ and use the dispersion relation (\ref{eq:ds1}) to obtain
\begin{equation}
   {\rm cn}(\bar x,-1){\rm dn}(\bar x,-1)w''-4{\rm sn}^3(\bar x,-1)w'=0
\end{equation}
where the primes mean derivative with respect to $\bar x$. From eq.(\ref{eq:sol1}) we can obtain the solution in the rest reference frame $p_1=p_2=0$ and $p_0=Ng^2/2^\frac{1}{4}$. The corresponding Green function is
\begin{equation}
    G_R(t)=-\frac{1}{\mu_0 2^\frac{3}{4}}\theta(t){\rm cn}(\mu_0 t +\phi,-1){\rm dn}(\mu_0 t+\phi,-1)
\end{equation}
where we have set $\mu_0=Ng^2/2^\frac{1}{4}$ for a reference mass and $\theta(t)$ the Heaviside function granting that the solution is different from 0 at $t>0$ and 0 for $t<0$ and provided that ${\rm cn}(\phi,-1)=0$. Similarly, one can define a backward propagating Green function as
\begin{equation}
    G_A(t)=\theta(-t){\rm cn}(-\mu_0 t +\phi,-1){\rm dn}(-\mu_0 t+\phi,-1).
\end{equation}
So, the propagator is
\begin{equation}
    G(t,0)=\delta^{d-1}(x)\left[G_A(t)+G_R(t)\right].
\end{equation}
When we turn to a Fourier transform, Fourier series of Jacobi functions are well-known \cite{NIST} and so one arrives, back to the moving reference frame, at the result
\begin{equation}
\label{eq:prop}
   G(p)=\sum_{n=0}^\infty\frac{B_n}{p^2-m_n^2+i\epsilon}
\end{equation}
with
\begin{equation}
    B_n=(2n+1)^2\frac{\pi^3}{4K^3(-1)}\frac{e^{-(n+\frac{1}{2})\pi}}{1+e^{-(2n+1)\pi}}.
\end{equation}
being $K(-1)$ the complete elliptic integral of the first kind and we get the mass spectrum
\begin{equation}
\label{eq:ms}
   m_n=(2n+1)\frac{\pi}{2K(-1)}\mu_0.
\end{equation}
So, our final result for the Green function in $d=2+1$ is
\begin{equation}
    G_{\mu\nu}^{ab}(p)=\delta_{ab}\left(g_{\mu\nu}-\frac{p_\mu p_\nu}{p^2}\right)G(p).
\end{equation}
This result implies that the Yang-Mills theory shows up a mass gap also in this case. The corresponding spectrum can be used to fit with lattice data.

\section{Next-to-leading order correction}
\label{sec3}

In order to evaluate the next-to-leading order correction, we use the technique outlined in \cite{Ramond:1981pw}. We have noted in Sec.\ref{sec2} that the Yang-Mills theory just reduces to the solution of the classical equation of the Green function for a scalar field and that this correspondence becomes exact in the Landau gauge. So, our approach will be to evaluate the next-to-leading order correction with the effective potential for the scalar field. Then, we will use such a correction to evaluate the corresponding correction in the Yang-Mills theory. The idea is to show that this correction becomes increasingly negligible at increasing 't~Hooft coupling.

\subsection{Scalar field theory}

Given the functional of the scalar theory for the ground state in (\ref{eq:exsol}) as in Ref.\cite{Frasca:2013tma}, we move it to $d=2+1$ noting that $\lambda=Ng^2$ has now the dimension of mass and the field $\phi$ that of the square root of a mass. Working in dimensionless unit, $y=\lambda x$, $\bar\phi=\phi/\sqrt{\lambda}$, $\bar j=\lambda^{-5/2}j$ and introducing the identity $\bar\phi=\bar\phi_0+\delta\phi$, one gets
\begin{equation}
  Z[j]={\cal N}e^{i\int d^3y\bar j(y)\bar\chi_{d=2+1}(y)}
  e^{-i\int d^3y\left(-\bar\chi_{d=2+1}(y)
	\frac{\delta^3}{i\delta\bar j(y)^3}+\frac{1}{4}\frac{\delta^4}{\delta\bar j(y)^4}\right)}
  e^{\frac{i}{2}\int d^3y'd^3y''\bar j(y')G(y'-y'')\bar j(y'')}.
\end{equation}
The bar over the variables means that these have been made dimensionless while the letter $y$ is used for dimensionless space-time variables. In the following we will set $\phi_0(x)=\chi_{d=2+1}(x)$ from eq.(\ref{eq:exsol}). This gives the strong coupling expansion for the scalar field in $d=2+1$ as already shown in \cite{Frasca:2013tma} for 3+1 dimensions. Powers in the functional derivative of the current determine the order of magnitude in the series expansion. This means that the quartic derivative is negligible with respect to the cubic one. In the constant ${\cal N}$ we have included the value of the action for $\phi_0$.

Now, we follow the procedure outlined in \cite{Ramond:1981pw}. Let us note that
\begin{eqnarray}
    \frac{\delta Z[\bar j]}{\delta\bar j(y)}&=&i\left[\bar\phi_0(y)+\right. \nonumber \\
		&&\left.e^{-i\int d^3y\left(-\bar\phi_0(y)
	\frac{\delta^3}{i\delta\bar j(y)^3}+\frac{1}{4}\frac{\delta^4}{\delta\bar j(y)^4}\right)}
	\int d^3y''G(y-y'')\bar j(y'')
	e^{i\int d^3y\left(-\bar\phi_0(y)
	\frac{\delta^3}{i\delta\bar j(y)^3}+\frac{1}{4}\frac{\delta^4}{\delta\bar j(y)^4}\right)}\right]
	\times \nonumber \\
	Z[\bar j]
\end{eqnarray}
and so, one has
\begin{eqnarray}
     (\partial^2+3\bar\phi_0^2(y))\frac{\delta Z[\bar j]}{\delta\bar j(y)}&=&
		i(\partial^2+3\bar\phi_0^2(y))\bar\phi_0(y)Z[\bar j]+\nonumber \\
		&&ie^{-i\int d^3y'\left(-\bar\phi_0(y)
	\frac{\delta^3}{i\delta\bar j(y')^3}+\frac{1}{4}\frac{\delta^4}{\delta\bar j(y')^4}\right)}
	\bar j(y)
	e^{i\int d^3y\left(-\bar\phi_0(y')
	\frac{\delta^3}{i\delta\bar j(y')^3}+\frac{1}{4}\frac{\delta^4}{\delta\bar j(y')^4}\right)}Z[\bar j].
\end{eqnarray}
Then
\begin{eqnarray}
     (\partial^2+3\bar\phi_0^2(y))\frac{\delta Z[\bar j]}{\delta\bar j(y)}&=&
		2i\bar\phi_0^3(y)Z[\bar j]+\nonumber \\
		&&i\left[\bar j(y)-
		\left(3\bar\phi_0(y)
	\frac{\delta^2}{\delta\bar j(y)^2}-\frac{\delta^3}{i\delta\bar j(y)^3}\right)\right]Z[\bar j].
\end{eqnarray}
where use has been made of the equation $(\partial^2+\bar\phi_0^2(y))\bar\phi_0(y)=0$. Finally, 
\begin{eqnarray}
     (\partial^2+3\bar\phi_0^2(y))\bar\phi_{cl}(y)&=&
		\bar j(y)+\nonumber \\
		&&2\bar\phi_0^3(y)-
		Z^{-i}[j]\left(3\bar\phi_0(y)
	\frac{\delta^2}{\delta\bar j(y)^2}-\frac{\delta^3}{i\delta\bar j(y)^3}\right)Z[\bar j]
\end{eqnarray}
having used the definition
\begin{equation}
   \bar\phi_{cl}(y)=-iZ^{-1}[j]\frac{\delta Z[\bar j]}{\delta\bar j(y)}.
\end{equation}
Then,
\begin{eqnarray}
     (\partial^2+3\bar\phi_0^2(y))\bar\phi_{cl}(y)&=&
		\bar j(y)+\nonumber \\
		&&2\bar\phi_0^3(y)-3i\bar\phi_0(y)\frac{\delta\bar\phi_{cl}(y)}{\delta\bar j(y)}+3\bar\phi_0(y)\bar\phi_{cl}^2(y)
		-\bar\phi_{cl}^3(y)+ \nonumber \\
		&&\frac{3}{2}i\frac{\delta\bar\phi_{cl}^2(y)}{\delta\bar j(y)}
		+\frac{\delta^2\bar\phi_{cl}(y)}{\delta\bar j(y)^2}
\end{eqnarray}
that is the equation of motion for $\bar\phi_{cl}$. This equation can be easily transformed into an integral equation as we know the Green function. This yields
\begin{eqnarray}
     \bar\phi_{cl}(y)&=&2\int d^3y'G(y-y')\bar\phi_0^3(y')+\int d^3y'G(y-y')\bar j(y') \nonumber \\
		&&-3i\int d^3y'G(y-y')\bar\phi_0(y')\frac{\delta\bar\phi_{cl}(y')}{\delta\bar j(y')}
		+3\int d^3y'G(y-y')\bar\phi_0(y')\bar\phi_{cl}^2(y') \nonumber \\
		&&-\int d^3y'G(y-y')\bar\phi_{cl}^3(y')+
		\frac{3}{2}i\int d^3y'G(y-y')\frac{\delta\bar\phi_{cl}^2(y')}{\delta\bar j(y')} \nonumber \\
		&&+\int d^3y'G(y-y')\frac{\delta^2\bar\phi_{cl}(y')}{\delta\bar j(y')^2}.
\end{eqnarray}
From this equation, containing also the quantum corrections, we can evaluate the next-to-leading order correction by starting to iterate with the terms $\bar\phi_{cl}^{(0)}(y)=2\int d^3y'G(y-y')\bar\phi_0^3(y')+\int d^3y'G(y-y')\bar j(y')$. Our aim is to obtain the one loop correction to the propagator in order to evaluate the Wilson loop. Then, the term linear in the current $\bar j$ will be
\begin{eqnarray}
     \bar\phi_{cl}(y)&=&\int d^3y'G(y-y')\bar j(y') \nonumber \\
		&&+12\int d^3y'G(y-y')\int d^3y''G(y'-y'')\bar\phi_0^3(y'')\int d^3y'''G(y'-y''')\bar j(y''') \nonumber \\
		&&-6\int d^3y'G(y-y')\left[\int d^3y''G(y'-y'')\bar\phi_0^3(y'')\right]^2\int d^3y'''G(y'-y''')\bar j(y''') \nonumber \\
		&&+3iG(0)\int d^3y'G(y-y')\int d^3y''G(y'-y'')\bar j(y'')+\ldots.
\end{eqnarray}
From this one gets immediately
\begin{eqnarray}
\label{eq:Gren}
     G_r(y,\bar y)=\left.\frac{\delta\bar\phi_{cl}(y)}{\delta\bar j(\bar y)}\right|_{\bar j=0}&=&G(y-\bar y) \nonumber \\
		&&+12\int d^3y'G(y-y')\int d^3y''G(y'-y'')\bar\phi_0^3(y'')G(y'-\bar y) \nonumber \\
		&&-6\int d^3y'G(y-y')\left[\int d^3y''G(y'-y'')\bar\phi_0^3(y'')\right]^2G(y'-\bar y) \nonumber \\
		&&+3iG(0)G(y-\bar y)+\ldots.
\end{eqnarray}
for the corrected Green function and use has been made of the identity $G(y-\bar y)=\int d^3y'G(y-y')G(y'-\bar y)$. We note the imaginary part that accounts for the finite width of the states. For the moment we do not consider it further. The interesting part of this expression is the real one that permits us to evaluate the wanted correction.

\subsection{Yang-Mills theory}

Now, we are able to evaluate the next-to-leading order correction to the gluon propagator just setting
\begin{equation}
    \Delta_{\mu\nu}^{ab}(p)=\delta_{ab}\left(g_{\mu\nu}-\frac{p_\mu p_\nu}{p^2}\right)G_r(p).
\end{equation}
We now evaluate the real part of the correction given in (\ref{eq:Gren}). Then
%In order to do this, we do the following
%\begin{equation}
%    \bar\phi_0(x)\rightarrow\bar\phi_0(0)
%\end{equation}
%ignoring the oscillations around the vacuum expectation value of the field. This is rather usual in quantum field theory. Then, the choice of the phase is determined by our choice of the propagator that granted ${\rm sn}(\phi,-1)=1$. So, $\bar\phi_0(0)=2^\frac{1}{4}$ in dimensionless units. This will yield
\begin{eqnarray}
\label{eq:Gren1}
    G_r(y,\bar y)&=&G(y-\bar y) \nonumber \\
		&&+2^\frac{3}{4}12\int d^3y'G(y-y')\int d^3y''G(y'-y'')\bar\phi_0^3(y'')G(y'-\bar y) \nonumber \\
		&&-2^\frac{1}{2}12\int d^3y'G(y-y')\left[\int d^3y''G(y'-y'')\bar\phi_0^3(y'')\right]^2G(y'-\bar y) \nonumber \\
		&&+\ldots.
\end{eqnarray}
We just note that
\begin{equation}
    {\rm sn}^3(z,-1)=\frac{\pi^3}{4K^3(-1)}\sum_{n=0}^\infty(-1)^n(2n+1)^2\frac{e^{-\left(n+\frac{1}{2}\right)\pi}}{1+e^{-(2n+1)\pi}}
		\sin\left(\frac{(2n+1)\pi z}{2K(-1)}\right)
\end{equation}
and so, $\bar\phi_0^3(y)$ has a Fourier series in closed form. Moving to momenta one has
\begin{eqnarray}
\label{eq:Gr}
    G_r(p)(2\pi)^3\delta^3(\bar p-\bar p')&=&G(p)(2\pi)^3\delta^3(\bar p-\bar p') \nonumber \\
		&&+2^\frac{3}{4}12[G(\bar p)]^2G(\bar p-\bar p')[\bar\phi_0^3](\bar p-\bar p') \nonumber \\
		&&-2^\frac{1}{2}12[G(\bar p)]^2\int\frac{d^3\bar p''}{(2\pi)^3}G(\bar p'')
		G(\bar p -\bar p' -\bar p'')[\bar\phi_0^3](\bar p'')[\bar\phi_0^3](\bar p-\bar p'-\bar p'') \nonumber \\
		&&+\ldots.
\end{eqnarray}
where $[\ldots]$ means Fourier transformed. It is not difficult to see that
\begin{equation}
   \bar\phi_0^3(x)=2^\frac{3}{4}\sqrt{(Ng^2)^3}\frac{\pi^3}{4K^3(-1)}\sum_{n=0}^\infty(2n+1)^2\frac{e^{-\left(n+\frac{1}{2}\right)\pi}}{1+e^{-(2n+1)\pi}}
		\cos\left(k_n\cdot x\right)
\end{equation}
being $k_n=\frac{(2n+1)\pi}{2K(-1)}k$ and noting that $\phi=(4m+1)K(-1)$ for $m=0,1,2,\ldots$ \cite{Frasca:2013tma}. We choose $m=0$ for the sake of simplicity. So, the Fourier transform is yielded by
\begin{equation}
   [\bar\phi_0^3](\bar p'')=(2\pi)^32^\frac{3}{4}\sqrt{(Ng^2)^3}\frac{\pi^3}{8K^3(-1)}\sum_{n=0}^\infty(2n+1)^2\frac{e^{-\left(n+\frac{1}{2}\right)\pi}}{1+e^{-(2n+1)\pi}}
	 [\delta^3(p-k_n)+\delta^3(p+k_n)]
\end{equation}
and we see that the propagator gets contributions from processes involving an arbitrary number of gluon excited states. We are interested in the contributions involving no gluon quanta at all. This implies immediately that the second term on the rhs of eq.(\ref{eq:Gr}) does not contain such terms. So, we consider
\begin{eqnarray}
    -2^\frac{1}{2}12[G(\bar p)]^2\int\frac{d^3\bar p''}{(2\pi)^3}G(\bar p'')
		G(\bar p -\bar p' -\bar p'')[\bar\phi_0^3](\bar p'')[\bar\phi_0^3](\bar p-\bar p'-\bar p'')&=& \nonumber \\
		-2^\frac{1}{2}12[G(\bar p)]^22^\frac{3}{4}\sqrt{(Ng^2)^3}\frac{\pi^3}{8K^3(-1)}\sum_{n=0}^\infty
		C_nG(k_n)G(\bar p -\bar p'-k_n)[\bar\phi_0^3](\bar p-\bar p'-k_n) &+& \nonumber \\
		-2^\frac{1}{2}12[G(\bar p)]^22^\frac{3}{4}\sqrt{(Ng^2)^3}\frac{\pi^3}{8K^3(-1)}\sum_{n=0}^\infty
		C_nG(-k_n)G(\bar p -\bar p'+k_n)[\bar\phi_0^3](\bar p-\bar p'+k_n) &=& \nonumber \\
		-2^\frac{1}{2}12(2\pi)^3[G(\bar p)]^22^\frac{3}{2}(Ng^2)^3\frac{\pi^6}{64K^6(-1)}\sum_{l,n=0}^\infty
		C_lC_n\left[G(k_n)G(k_l+k_n)\delta^3(\bar p -\bar p'-k_l-k_n)\right.&+& \nonumber \\
		\left.G(k_n)G(-k_l+k_n)\delta^3(\bar p -\bar p'+k_l-k_n)\right] &+& \nonumber \\
		-2^\frac{1}{2}12(2\pi)^3[G(\bar p)]^22^\frac{3}{2}(Ng^2)^3\frac{\pi^6}{64K^6(-1)}\sum_{l,n=0}^\infty
		C_lC_n\left[G(-k_n)G(k_l-k_n)\delta^3(\bar p -\bar p'-k_l+k_n)\right.&+& \nonumber \\
		\left.G(-k_n)G(-k_l-k_n)\delta^3(\bar p -\bar p'+k_l+k_n)\right]
\end{eqnarray}
where we have set $C_n=(2n+1)^2\frac{e^{-\left(n+\frac{1}{2}\right)\pi}}{1+e^{-(2n+1)\pi}}$. The terms we are interested to are those with $n=l$ and noting that $G(-k)=G(k)$ we get the contribution at this order
\begin{equation}
-(2\pi)^3\delta^3(\bar p -\bar p')[G(\bar p)]^2(Ng^2)^3\frac{3\pi^6}{K^6(-1)}\sum_{n=0}^\infty
		C_n^2G(k_n)G(0)
\end{equation}
and so
\begin{eqnarray}
\label{eq:Gr1}
    G_r(p)(2\pi)^3\delta^3(\bar p-\bar p')&=&G(p)(2\pi)^3\delta^3(\bar p-\bar p')\times \nonumber \\
		&&\left[1-G(p)(Ng^2)^3\frac{3\pi^6}{K^6(-1)}\sum_{n=0}^\infty
		C_n^2G(k_n)G(0)+\ldots\right]
\end{eqnarray}
where we have omitted the bar over momenta. Now,
%\begin{equation}
%    G(k_n)=G(0)+\sum_{m\ne n}\frac{B_n}{m_0^2}\frac{1}{(2n+1)^2-(2m+1)^2}
%\end{equation}
it is not difficult to verify that the higher term goes like $(Ng^2)^3G(k_n)G(0)\sim (Ng^2)^3/\mu_0^4\propto 1/Ng^2$, modulo renormalization, and so we can omit it in the discussion for the computation of the ground state. Indeed, this adds a term negligible at increasing 't~Hooft coupling as we guessed at the start. As a final note we point out the peculiar singularity of $G(k_n)$ when we work on-shell with the dispersion relation (\ref{eq:ds2+1}). We will not discuss this matter here being of no interest for the main results of the paper.
     
\section{Ground state}
\label{sec4}

From eq.(\ref{eq:ms}) we can perform a comparison with lattice data \cite{Teper:1998te,Lucini:2002wg,Bringoltz:2006zg}. The result given in \cite{Lucini:2002wg,Bringoltz:2006zg} for the ratio of string tension $\sqrt{\sigma}$ to 't~Hooft coupling $Ng^2$ is
\begin{equation}
\label{eq:tl}
    \frac{\sqrt{\sigma}}{Ng^2}=0.19755(34)-\frac{0.1200(29)}{N^2}.
\end{equation}
We compare eq.(\ref{eq:ms}) with lattice results presented in \cite{Lucini:2002wg} for $N\rightarrow\infty$ that are those consistent with our computations. 
%So, in this case, it is a good approximation to take $Ng^2/\sqrt{\sigma}\approx 5.06$.
Firstly we consider the result for the string tension in Nair and Karabali \cite{Nair:2002yg,Karabali:2009rg} that, even if not perfectly aligned with lattice results \cite{Lucini:2002wg,Bringoltz:2006zg}, is an astonishingly good approximation
\begin{equation}
    \sqrt{\sigma}=g^2\sqrt{\frac{N^2-1}{8\pi}}.
\end{equation}
This result was neatly improved in \cite{Karabali:2009rg}. Then,
\begin{equation}
    \frac{m_n}{\sqrt{\sigma}}=(2n+1)\frac{2^\frac{1}{4}\pi^\frac{3}{2}}{K(-1)}\frac{N}{\sqrt{N^2-1}}\approx
		(2n+1)\frac{2^\frac{1}{4}\pi^\frac{3}{2}}{K(-1)}\left[1+\frac{1}{2N^2}++\frac{3}{8N^4}+O(1/N^6)\right].
\end{equation}
We identify
\begin{equation}
   \frac{\sqrt{\sigma}}{Ng^2}=\frac{K(-1)}{2^\frac{1}{4}\pi^\frac{3}{2}}\left[1-\frac{1}{2N^2}-\frac{1}{8N^4}+O(1/N^6)\right]
\end{equation}
that yields
\begin{equation}
\label{eq:Ng2oversigma}
   \frac{\sqrt{\sigma}}{Ng^2}=0.1979839190\ldots-\frac{0.09899195950\ldots}{N^2}-\frac{0.02474798988\ldots}{N^4}+O(1/N^6)
\end{equation}
where we used dots to remember we are working with pure numbers. We see that the agreement with lattice result (\ref{eq:tl}) is exceedingly good. So, introducing the string tension computed by Nair and Karabali into our equation for the ratio $\frac{\sqrt{\sigma}}{Ng^2}$ produced an astonishingly good agreement with lattice data in the limit $N\rightarrow\infty$.

\begin{table}[H]
\begin{center}
\begin{tabular}{|c|c|c|c|} \hline\hline
Order & Lattice     & Theoretical & Error    \\ \hline
0   & 0.19755(34) & 0.19798     & 0.2\%    \\ \hline 
1   & 0.1200(29)  & 0.09899     & 17\%     \\ \hline
2   & -           & 0.02474     & .        \\ \hline\hline
\end{tabular}
\caption{\label{tab:cc} Comparison for coefficients of the ratio of string tension to 't~Hooft coupling at large $N$.}
\end{center}
\end{table}

We note that the combination of our formula with that by Karabali and Nair aligns the final value toward the lattice data at the leading order.

For the analysis of the ground state it is required a corrective factor of $\sqrt{2/3}$ instead. A similar situation was discussed in \cite{Johnson:2000qz} for the Isgur-Paton model \cite{Isgur:1984bm} and agrees rather well with the idea of flux tubes in hadrons \cite{Brodsky:2014yha} supporting the results of AdS/CFT. Indeed, the two factors, ours and that computed by Johnson and Teper \cite{Johnson:2000qz}, are the same. We get
\begin{equation}
    \frac{m_{0^{++}}}{\sqrt{\sigma}}=\frac{\pi}{2^\frac{5}{4}K(-1)}\left[\frac{Ng^2}{\sqrt{\sigma}}\right]\sqrt{\frac{2}{3}}=
		\frac{2^\frac{1}{4}\pi^\frac{3}{2}}{K(-1)}\left[1+\frac{1}{2N^2}+\frac{3}{8N^4}+O(1/N^6)\right]\sqrt{\frac{2}{3}}.
\end{equation}
and we use the result (\ref{eq:Ng2oversigma}) we obtain the results in Tab.~\ref{tab:0++} that are in exceedingly good agreement with lattice computations.

\begin{table}[H]
\begin{center}
\begin{tabular}{|c|c|c|c|} \hline\hline
$n$ & Lattice   & Theoretical & Error \\ \hline
0   & 4.108(20) & 4.124055050 & 0.4\% \\ \hline 
1   & -         & 12.37216515 & -     \\ \hline\hline
\end{tabular}
\caption{\label{tab:0++} Comparison for the ground state and value of the next state for $N\rightarrow\infty$.}
\end{center}
\end{table}

This result improves neatly with respect to the works \cite{Leigh:2005dg,Leigh:2006vg} for the ground state of the theory.

\section{Conclusions}
\label{sec5}

We developed a different approach with respect to Nair and Karabali to study Yang-Mills theory in 2+1 dimensions. Our aim was to validate the results of these authors using a different technique. The combined result of the mass gap we compute and the string tension obtained by Nair and Karabali shows an agreement with lattice data that is neatly improved at the leading order or $N\rightarrow\infty$.

Considering the ground state of the theory, we are able to confirm the conclusions by Teper and Johnson about Isgur and Paton model that describes hadrons using flux tubes. The ground state is recovered with a correction factor of $\sqrt{2/3}$ yielding a exceedingly good agreement with lattice data and improving with respect to the work by Leigh, Minic and Yelnikov for the mass gap. All in all, our technique grants a strong confirmation to Karabali and Nair approach.


\begin{thebibliography}{99}
%\cite{Durr:2008zz}
\bibitem{Durr:2008zz} 
  S.~Durr, Z.~Fodor, J.~Frison, C.~Hoelbling, R.~Hoffmann, S.~D.~Katz, S.~Krieg and T.~Kurth {\it et al.},
  %``Ab-Initio Determination of Light Hadron Masses,''
  Science {\bf 322}, 1224 (2008)
  [arXiv:0906.3599 [hep-lat]].
  %%CITATION = ARXIV:0906.3599;%%
  %304 citations counted in INSPIRE as of 19 Aug 2014
	
%\cite{Bazavov:2009bb}
\bibitem{Bazavov:2009bb} 
  A.~Bazavov, D.~Toussaint, C.~Bernard, J.~Laiho, C.~DeTar, L.~Levkova, M.~B.~Oktay and S.~Gottlieb {\it et al.},
  %``Nonperturbative QCD simulations with 2+1 flavors of improved staggered quarks,''
  Rev.\ Mod.\ Phys.\  {\bf 82}, 1349 (2010)
  [arXiv:0903.3598 [hep-lat]].
  %%CITATION = ARXIV:0903.3598;%%
  %251 citations counted in INSPIRE as of 19 Aug 2014
	
%\cite{Maas:2011se}
\bibitem{Maas:2011se} 
  A.~Maas,
  %``Describing gauge bosons at zero and finite temperature,''
  Phys.\ Rept.\  {\bf 524}, 203 (2013)
  [arXiv:1106.3942 [hep-ph]].
  %%CITATION = ARXIV:1106.3942;%%
  %55 citations counted in INSPIRE as of 19 Aug 2014
	
%\cite{Petreczky:2012rq}
\bibitem{Petreczky:2012rq} 
  P.~Petreczky,
  %``Lattice QCD at non-zero temperature,''
  J.\ Phys.\ G {\bf 39}, 093002 (2012)
  [arXiv:1203.5320 [hep-lat]].
  %%CITATION = ARXIV:1203.5320;%%
  %53 citations counted in INSPIRE as of 19 Aug 2014
	
%\cite{Teper:1998te}
\bibitem{Teper:1998te} 
  M.~J.~Teper,
  %``SU(N) gauge theories in (2+1)-dimensions,''
  Phys.\ Rev.\ D {\bf 59}, 014512 (1999)
  [hep-lat/9804008].
  %%CITATION = HEP-LAT/9804008;%%
  %249 citations counted in INSPIRE as of 28 Jul 2014

%\cite{Lucini:2002wg}
\bibitem{Lucini:2002wg} 
  B.~Lucini and M.~Teper,
  %``SU(N) gauge theories in (2+1)-dimensions: Further results,''
  Phys.\ Rev.\ D {\bf 66}, 097502 (2002)
  [hep-lat/0206027].
  %%CITATION = HEP-LAT/0206027;%%
  %78 citations counted in INSPIRE as of 28 Jul 2014
	
%\cite{Bringoltz:2006zg}
\bibitem{Bringoltz:2006zg} 
  B.~Bringoltz and M.~Teper,
  %``A Precise calculation of the fundamental string tension in SU(N) gauge theories in 2+1 dimensions,''
  Phys.\ Lett.\ B {\bf 645}, 383 (2007)
  [hep-th/0611286].
  %%CITATION = HEP-TH/0611286;%%
  %43 citations counted in INSPIRE as of 18 Aug 2014
	
%\cite{Caselle:2011fy}
\bibitem{Caselle:2011fy} 
  M.~Caselle, L.~Castagnini, A.~Feo, F.~Gliozzi and M.~Panero,
  %``Thermodynamics of SU(N) Yang-Mills theories in 2+1 dimensions I - The confining phase,''
  JHEP {\bf 1106}, 142 (2011)
  [arXiv:1105.0359 [hep-lat]].
  %%CITATION = ARXIV:1105.0359;%%
  %22 citations counted in INSPIRE as of 21 Aug 2014
	
%\cite{Caselle:2011mn}
\bibitem{Caselle:2011mn} 
  M.~Caselle, L.~Castagnini, A.~Feo, F.~Gliozzi, U.~Gursoy, M.~Panero and A.~Schafer,
  %``Thermodynamics of SU(N) Yang-Mills theories in 2+1 dimensions II. The Deconfined phase,''
  JHEP {\bf 1205}, 135 (2012)
  [arXiv:1111.0580 [hep-th]].
  %%CITATION = ARXIV:1111.0580;%%
  %23 citations counted in INSPIRE as of 21 Aug 2014
	
%\cite{Nair:2002yg}
\bibitem{Nair:2002yg} 
  V.~P.~Nair,
  %``Yang-Mills theory in (2+1)-dimensions: A Short review,''
  Nucl.\ Phys.\ Proc.\ Suppl.\  {\bf 108}, 194 (2002)
  [hep-th/0204063] 
	and Refs. therein.
  %%CITATION = HEP-TH/0204063;%%
  %8 citations counted in INSPIRE as of 18 Aug 2014

%\cite{Karabali:2009rg}
\bibitem{Karabali:2009rg} 
  D.~Karabali, V.~P.~Nair and A.~Yelnikov,
  %``The Hamiltonian Approach to Yang-Mills (2+1): An Expansion Scheme and Corrections to String Tension,''
  Nucl.\ Phys.\ B {\bf 824}, 387 (2010)
  [arXiv:0906.0783 [hep-th]].
  %%CITATION = ARXIV:0906.0783;%%
  %24 citations counted in INSPIRE as of 18 Aug 2014
	
%\cite{Leigh:2005dg}
\bibitem{Leigh:2005dg} 
  R.~G.~Leigh, D.~Minic and A.~Yelnikov,
  %``Solving pure QCD in 2+1 dimensions,''
  Phys.\ Rev.\ Lett.\  {\bf 96}, 222001 (2006)
  [hep-th/0512111].
  %%CITATION = HEP-TH/0512111;%%
  %30 citations counted in INSPIRE as of 19 Aug 2014
	
%\cite{Leigh:2006vg}
\bibitem{Leigh:2006vg} 
  R.~G.~Leigh, D.~Minic and A.~Yelnikov,
  %``On the Glueball Spectrum of Pure Yang-Mills Theory in 2+1 Dimensions,''
  Phys.\ Rev.\ D {\bf 76}, 065018 (2007)
  [hep-th/0604060].
  %%CITATION = HEP-TH/0604060;%%
  %31 citations counted in INSPIRE as of 19 Aug 2014
	
%\cite{Karabali:1995ps}
\bibitem{Karabali:1995ps} 
  D.~Karabali and V.~P.~Nair,
  %``A Gauge invariant Hamiltonian analysis for nonAbelian gauge theories in (2+1)-dimensions,''
  Nucl.\ Phys.\ B {\bf 464}, 135 (1996)
  [hep-th/9510157].
  %%CITATION = HEP-TH/9510157;%%
  %91 citations counted in INSPIRE as of 19 Aug 2014
	
%\cite{Frasca:2013tma}
\bibitem{Frasca:2013tma} 
  M.~Frasca,
  %``Scalar field theory in the strong self-interaction limit,''
  Eur.\ Phys.\ J.\ C {\bf 74}, 2929 (2014)
  [arXiv:1306.6530 [hep-ph]].
  %%CITATION = ARXIV:1306.6530;%%
  %2 citations counted in INSPIRE as of 24 Jul 2014
	
%\cite{Frasca:2009bc}
\bibitem{Frasca:2009bc} 
  M.~Frasca,
  %``Exact solutions of classical scalar field equations,''
  J.\ Nonlin.\ Math.\ Phys.\  {\bf 18}, 291 (2011)
  [arXiv:0907.4053 [math-ph]].
  %%CITATION = ARXIV:0907.4053;%%
  %5 citations counted in INSPIRE as of 17 Jun 2013%\cite{Nair:2005iw}

%\cite{Frasca:2009yp}
\bibitem{Frasca:2009yp} 
  M.~Frasca,
  %``Mapping a Massless Scalar Field Theory on a Yang-Mills Theory: Classical Case,''
  Mod.\ Phys.\ Lett.\ A {\bf 24}, 2425 (2009)
  [arXiv:0903.2357 [math-ph]].
  %%CITATION = ARXIV:0903.2357;%%
  %21 citations counted in INSPIRE as of 24 Jul 2014
	
%\cite{Brodsky:2014yha}
\bibitem{Brodsky:2014yha} 
  S.~J.~Brodsky, G.~F.~de Teramond, H.~G.~Dosch and J.~Erlich,
  %``Light-Front Holographic QCD and Emerging Confinement,''
  arXiv:1407.8131 [hep-ph].
  %%CITATION = ARXIV:1407.8131;%%
  %1 citations counted in INSPIRE as of 19 Aug 2014
	
%\cite{Isgur:1984bm}
\bibitem{Isgur:1984bm} 
  N.~Isgur and J.~E.~Paton,
  %``A Flux Tube Model for Hadrons in QCD,''
  Phys.\ Rev.\ D {\bf 31}, 2910 (1985).
  %%CITATION = PHRVA,D31,2910;%%
  %585 citations counted in INSPIRE as of 18 Aug 2014
		
%\cite{Johnson:2000qz}
\bibitem{Johnson:2000qz} 
  R.~W.~Johnson and M.~J.~Teper,
  %``String models of glueballs and the spectrum of SU(N) gauge theories in (2+1)-dimensions,''
  Phys.\ Rev.\ D {\bf 66}, 036006 (2002)
  [hep-ph/0012287].
  %%CITATION = HEP-PH/0012287;%%
  %27 citations counted in INSPIRE as of 18 Aug 2014
	
%\cite{Cahill:1985mh}
\bibitem{Cahill:1985mh} 
  R.~T.~Cahill and C.~D.~Roberts,
  %``Soliton Bag Models of Hadrons from QCD,''
  Phys.\ Rev.\ D {\bf 32}, 2419 (1985).
  %%CITATION = PHRVA,D32,2419;%%
  %203 citations counted in INSPIRE as of 25 Jul 2014
	
%\cite{Rubakov:2002fi}
\bibitem{Rubakov:2002fi} 
  V.~A.~Rubakov,
  ``Classical theory of gauge fields'',
  Princeton, USA: Univ. Pr. (2002), pp.75ff.
	
\bibitem{NIST}
F.~W.~J.~Olver, D.~W.~Lozier, R.~F.~Boisvert, C.~W.~Clark (Eds.),
``NIST Handbook of Mathematical Functions'',
Cambridge, UK: Univ. Pr. (2010), Ch. 22.
	
%\cite{Ramond:1981pw}
\bibitem{Ramond:1981pw} 
  P.~Ramond,
  %``Field Theory. A Modern Primer,''
  Front.\ Phys.\  {\bf 51}, 1 (1981), pp. 101ff.
  %%CITATION = FRPHA,51,1;%%
  %19 citations counted in INSPIRE as of 30 Jul 2014
	
\end{thebibliography}
\end{document}